\documentclass{aa}
\usepackage{graphicx}
\usepackage{txfonts}

\begin{document}

\title{NLTE strontium abundance in a sample of extremely metal poor stars and
the Sr/Ba ratio in the early Galaxy}

\author {
S.M. Andrievsky\inst{1,2}\and
F. Spite\inst{1}\and
S.A. Korotin\inst{2}\and
P. Fran\c cois\inst{1}\and
M. Spite\inst{1}\and
P. Bonifacio\inst{1,3}\and
R. Cayrel\inst{1}\and
V. Hill\inst{1}
}

\institute {
   GEPI, Observatoire de Paris-Meudon,  CNRS,  Universit\'e Paris Diderot, 
F-92125 Meudon Cedex France,
\and
   Department of Astronomy and Astronomical Observatory, Odessa
   National University, and Isaac Newton Institute of Chile Odessa branch,
   Shevchenko Park, 65014 Odessa, Ukraine, e-mail : {\tt scan@deneb1.odessa.ua}
   \and
   Istituto Nazionale di Astrofisica - Osservatorio Astronomico di Trieste, 
   via Tiepolo 11, 34143 Trieste, Italy
   }

\date{}

\authorrunning{Andrievsky et al.}
\titlerunning{NLTE strontium abundance}

\abstract
{Heavy element abundances in extremely metal-poor stars provide strong 
constraints on the processes of forming these elements in the first stars.}
{We attempt to determine   precise abundances of strontium in a homogeneous sample  
of extremely metal-poor stars.}
{The abundances of strontium in 54 very or extremely metal-poor 
stars, was redetermined by abandoning the local thermodynamic equilibrium (LTE)
hypothesis, and fitting non-LTE (NLTE) profiles to the observed spectral lines. 
The corrected Sr abundances and previously obtained NLTE Ba abundances are compared 
to the predictions of several hypothetical formation processes for the lighter 
neutron-capture elements.}
{Our NLTE abundances confirm the previously determined huge scatter of the strontium 
abundance in low metallicity stars. 
This scatter is also found (and is even larger) at very low metallicities (i. e. 
early in the chemical evolution). 
The Sr abundance in the extremely metal-poor (EMP) stars is compatible with 
the main r-process involved in other processes (or by variations of the 
r-process), as we briefly discuss.}
{}


\maketitle

\section{Introduction}   

Until recently, our knowledge of the properties of the first stellar 
generations of the Milky Way was quite restricted. Large telescopes 
and efficient detectors have made it possible to determine the abundances
of many chemical  elements of very primitive (very metal-poor) stars, 
constraining the theories of nucleosynthesis in the first stars.

The analysis of extremely metal-poor stars (EMP)  provide important 
information about the abundances of the elements produced by the first stars, 
and about the corresponding formation processes. The abundances of the heavy elements 
($Z>30$) are particularly interesting. Several processes have been proposed to explain 
their production, especially two important neutron-capture processes: the rapid process 
(r-process) and the slow process (s-process). Each of these processes is divided into 
two  sub-processes ("main" and "weak").

Massive stars with a very short lifetime  end their lives as core-collapse SN~II, 
forming neutron stars, and it had been assumed that the primitive massive 
stars provide a promising site for the early production of the 
neutron-capture elements by the r-process (see e.g. Cowan \& Thielemann 
\cite{CT04}, Farouqi et al. \cite{FKM}). 
The hypothesis was thought to be consistent with these massive 
stars being the primary early producers of light elements  
(called $\alpha$ elements), this production declining at later
phases of  the chemical evolution of the Galaxy, as (roughly) for the 
r-process elements.
 
Howeve, developments in nucleosynthesis models (e.g.  H\"udepold et al. \cite{HUD09}, 
Fisher et al. \cite{FIS09}, and Roberts et al. \cite{ROB10}) show that the classical 
neutrino-driven wind of core-collapse SNe is unable to produce the r-elements :
they could instead be produced independently in other ways as proposed by Wanajo et al. 
(\cite{WJ10}).

The neutron-capture elements may also be produced by  a "slow" neutron flux, 
but in the "main" s-process the production  takes place in intermediate-mass 
AGB stars, which have a long evolution time, that is inconsistent with an early production. 
Some early production could take place by means of the "weak" s-process, in massive 
primitive stars although the lack of seed nuclei places this idea in doubt.

In the Solar System, the global abundances of the heavy elements are the result of 
a late production by the main s-process in addition to an early production by 
the r-process,  and it has been possible to distinguish the fraction produced by 
each process.

Cowan \& Sneden (\cite{CS06}) demonstrated that the abundances of heavy 
($56 < Z < 70$) elements, display nearly identical abundance patterns
in the r-fraction of the abundances of the Solar System, and in 6 r-rich EMP 
stars (some more r-rich EMP stars have been added recently to the list). 
As a result, the main r-process has been proposed to be universal. With larger 
samples of analyzed stars, the whole picture appeared more complex. It is 
therefore quite important to determine the precise abundances of heavy elements 
in EMP stars. Since the electron concentration in the atmospheric plasma of 
these stars is substantially lower, significant deviations from LTE in these 
atmospheres are expected. Thus, reliable abundances of these stars 
can be obtained only by means of a NLTE analysis. As noted by Mashonkina \& Gehren 
(\cite{MG01}), the NLTE correction can be significant, and clearly differs 
from one star to another. We therefore decided to reanalyze the sample 
of the EMP stars previously studied by Fran\c cois et al. (\cite {FDH07}) and Bonifacio 
et al. (\cite{BSC09}) in the framework of the ESO large program "First stars, 
first nucleosynthesis". For this work, we determined for these stars the abundance 
of Sr based on NLTE computations of the line profiles. We combine these abundances 
with previous NLTE determinations of Ba abundances (see Andrievsky et al., \cite {ASK09}) 
to study the evolution of the [Sr/Ba] ratio in the early phases of the chemical evolution 
of the Galaxy.
  
\section{Sample of stars and their parameters}  

The spectra of the stars in this sample were presented in detail in 
Cayrel et al. (\cite{CDS04}) and Bonifacio et al. (\cite{BMS07}, \cite{BSC09}).   
About fifty stars (35 giants and 18 turnoff stars), most of them with $\rm [Fe/H] <-3$, 
were observed with the high-resolution spectrograph UVES (Dekker et al., \cite{DDK00}) 
fed by the ESO-VLT. The resolving power in the blue is $R \approx 45000$.  
The spectra were reduced using the UVES reduction tool within MIDAS (Ballester et
al., \cite{BMB00}). The signal-to-noise ratio (S/N) in the region of the blue 
strontium lines is typically $\sim 120$ per pixel with there being an average of 5 pixels 
per resolution element. The sample includes a few well-known typical stars for comparison.

The fundamental parameters ($\rm T_{eff}$, $\log~g$, [Fe/H]) of the stars 
were derived by Cayrel et al. (\cite{CDS04}) for the giants and Bonifacio et al. 
(\cite{BMS07}) for the turn-off stars. The temperature of the giants is deduced 
from the colour by adopting the calibration of Alonso et al. (\cite {AAM99}), 
and the temperature of the turn-off stars is deduced from the wings of the H$\alpha$ line. 
The gravities are derived from the ionisation equilibrium of iron (in LTE 
approximation), and could be affected by NLTE effects. These parameters are 
repeated in Table~\ref{tabstars}.

Briefly, MARCS model atmospheres have been used, but, for convenience, some 
parts of the computations have used ATLAS9 (Kurucz \cite{KurCD13}, \cite{K2005}) 
model atmospheres with the overshooting option. It has been shown (Castelli et al.  
\cite{CGK97}) that for EMP stars, non-overshooting models are more appropriate.  
Non-overshooting Kurucz models have been shown to provide abundances very similar  
(within 0.05 dex) to those of the MARCS models used by Cayrel et al. (\cite{CDS04})  
and Bonifacio et al. (\cite{BSC09}).

Table~\ref{tabstars} provides the adopted parameters and the Sr abundances. 
In addition we present the Ba abundances determined in the same way in Andrievsky et al. 
(\cite{ASK09}) but using Kurucz models with the overshooting option 
(Kurucz \cite{KurCD13}, \cite{K2005}). As a consequence,  these previous NLTE Ba abundances 
are shifted  (Table~\ref{tabstars}) by --0.03 dex for the turn-off stars, 
and --0.05 dex for the giants. 

Three new giant stars were added to the original sample, their parameters and 
NLTE abundances (including barium) being determined  in the same way; they 
are listed at the end of Table~\ref{tabstars}. In the last column, we indicate 
the number of lines of \ion{Sr}{ii} measured. 
 
Two carbon-rich stars (CEMPs) are included in the sample (they are designated 
by an asterisk in the first column of Table~\ref{tabstars}): CS 22949--037 
(Sr-rich) and CS  22892--052 (a useful typical example of an extremely r-rich 
star). Among the turn-off stars, CS 29527--015 is a spectroscopic binary.

\begin{table*}
\begin{center}    
\caption[]{Parameters of program stars and strontium abundance. The Ba abundance
has been corrected for overshooting.}
\label{tabstars}
\begin{tabular}{lcccccccccccc}
\hline
star      & T$_{\rm eff},\,K$ & $\log~g$ & V$_{\rm t}$, km~s$^{-1}$ &
[Fe/H] & $\epsilon$(Sr) &[Sr/H] & [Sr/Fe] & [Ba/H] &Rem\\
turnoff stars \\
\hline
BS~16023--046 & 6360 & 4.5 & 1.4 &  --2.97 & --0.08 & --3.00 &--0.03 & --    & 2\\
BS~16968--061 & 6040 & 3.8 & 1.5 &  --3.05 & --0.46 & --3.38 &--0.33 & --    & 2\\
BS~17570--063 & 6240 & 4.8 & 0.5 &  --2.92 &  +0.10 & --2.82 & +0.10 & --2.97& 2\\
CS~22177--009 & 6260 & 4.5 & 1.2 &  --3.10 & --0.09 & --3.01 & +0.09 & --    & 2\\
CS~22888--031 & 6150 & 5.0 & 0.5 &  --3.30 & --0.13 & --3.05 & +0.25 & --    & 2\\
CS~22948--093 & 6360 & 4.3 & 1.2 &  --3.30 & --0.20 & --3.12 & +0.18 & --2.97& 2\\
CS~22965--054 & 6090 & 3.8 & 1.4 &  --3.04 &  +0.20 & --2.72 & +0.32 & --    & 2\\
CS~22966--011 & 6200 & 4.8 & 1.1 &  --3.07 &  +0.07 & --2.85 & +0.22 & --2.87& 2\\
CS~29499--060 & 6320 & 4.0 & 1.5 &  --2.70 & --0.15 & --3.07 & -0.37 & --    & 2\\
CS~29506--007 & 6270 & 4.0 & 1.7 &  --2.91 &  +0.31 & --2.61 & +0.30 & --2.37& 2\\
CS~29506--090 & 6300 & 4.3 & 1.4 &  --2.83 &  +0.54 & --2.38 & +0.45 & --2.77& 2\\
CS~29518--043 & 6430 & 4.3 & 1.3 &  --3.20 &  +0.00 & --2.92 & +0.28 & --    & 2\\
CS~29527--015 & 6240 & 4.0 & 1.6 &  --3.55 &  -0.06 & --2.98 & +0.57 & --    & 2\\
CS~30301--024 & 6330 & 4.0 & 1.6 &  --2.75 &  +0.27 & --2.65 & +0.10 & --2.67& 2\\
CS~30339--069 & 6240 & 4.0 & 1.3 &  --3.08 & --0.11 & --3.03 & +0.05 & --    & 2\\
CS~31061--032 & 6410 & 4.3 & 1.4 &  --2.58 &  +0.55 & --2.37 & +0.21 & --    & 2\\
\hline
giants  \\
\hline
HD~2796        & 4950 & 1.5 & 2.1 &  --2.47 & +0.57 & --2.35 &  +0.12 & --2.68  & 4\\
HD~122563      & 4600 & 1.1 & 2.0 &  --2.82 & +0.10 & --2.82 &  +0.00 & --3.67  & 5\\
HD~186478      & 4700 & 1.3 & 2.0 &  --2.59 & +0.57 & --2.35 &  +0.24 & --2.67  & 4\\
BD~+17$\degr$3248    & 5250 & 1.4 & 1.5 &  --2.07 & +1.17 & --1.75 &  +0.32 & --1.74  & 2\\
BD~--18$\degr$5550   & 4750 & 1.4 & 1.8 &  --3.06 &--0.93 & --3.85 &  --0.79& --3.67  & 3\\
CD~--38$\degr$245    & 4800 & 1.5 & 2.2 &  --4.19 &--1.75 & --4.67 &  --0.48& --4.72  & 2\\
BS~16467--062  & 5200 & 2.5 & 1.6 &  --3.77 &--2.30 & --5.22 &  --1.45& --      & 2\\
BS~16477--003  & 4900 & 1.7 & 1.8 &  --3.36 &--0.40 & --3.32 &  +0.04 & --3.62  & 2\\
BS~17569--049  & 4700 & 1.2 & 1.9 &  --2.88 & +0.45 & --2.47 &  +0.41 & --2.72  & 4\\
CS~22169--035  & 4700 & 1.2 & 2.2 &  --3.04 &--0.33 & --3.25 &  --0.21& --4.17  & 2\\
CS~22172--002  & 4800 & 1.3 & 2.2 &  --3.86 &--2.03 & --4.95 &  --1.09& --4.82  & 2\\
CS~22186--025  & 4900 & 1.5 & 2.0 &  --3.00 &--0.18 & --3.10 &  --0.10& --2.94  & 2\\
CS~22189--009  & 4900 & 1.7 & 1.9 &  --3.49 &--1.28 & --4.20 &  --0.71& --4.62  & 2\\
CS~22873--055  & 4550 & 0.7 & 2.2 &  --2.99 &--0.40 & --3.32 &  --0.33& --3.37  & 2\\
CS~22873--166  & 4550 & 0.9 & 2.1 &  --2.97 & +0.27 & --2.65 &  +0.32 & --3.54  & 3\\
CS~22878--101  & 4800 & 1.3 & 2.0 &  --3.25 &--0.58 & --3.50 &  --0.25& --3.57  & 2\\
CS~22885--096  & 5050 & 2.6 & 1.8 &  --3.78 &--2.00 & --4.92 &  --1.14& --4.72  & 2\\
CS~22891--209  & 4700 & 1.0 & 2.1 &  --3.29 &--0.23 & --3.15 &  +0.14 & --3.65  & 2\\
CS~22892--052* & 4850 & 1.6 & 1.9 &  --3.03 & +0.47 & --2.45 &  +0.58 & --2.25  & 4\\
CS~22896--154  & 5250 & 2.7 & 1.2 &  --2.69 & +0.67 & --2.25 &  +0.44 & --2.35  & 3\\
CS~22897--008  & 4900 & 1.7 & 2.0 &  --3.41 &--0.08 & --3.00 &  +0.41 & --4.37  & 2\\
CS~22948--066  & 5100 & 1.8 & 2.0 &  --3.14 &--0.68 & --3.60 &  --0.46& --3.97  & 2\\
CS~22949--037* & 4900 & 1.5 & 1.8 &  --3.97 &--0.88 & --3.80 &  +0.17 & --4.47  & 2\\
CS~22952--015  & 4800 & 1.3 & 2.1 &  --3.43 &--1.33 & --4.25 &  --0.82& --4.57  & 2\\
CS~22953--003  & 5100 & 2.3 & 1.7 &  --2.84 & +0.30 & --2.62 &  +0.22 & --2.45  & 2\\
CS~22956--050  & 4900 & 1.7 & 1.8 &  --3.33 &--0.80 & --3.72 &  --0.39& --3.95  & 2\\
CS~22966--057  & 5300 & 2.2 & 1.4 &  --2.62 & +0.12 & --2.80 &  --0.18& --2.95  & 2\\
CS~22968--014  & 4850 & 1.7 & 1.9 &  --3.56 &--2.17 & --5.09 &  --1.53& --      & 2\\
CS~29491--053  & 4700 & 1.3 & 2.0 &  --3.04 &--0.33 & --3.25 &  --0.21& --3.74  & 2\\
CS~29495--041  & 4800 & 1.5 & 1.8 &  --2.82 &--0.13 & --3.05 &  --0.23& --3.28  & 2\\
CS~29502--042  & 5100 & 2.5 & 1.5 &  --3.19 &--1.98 & --4.90 &  --1.71& --4.62  & 2\\
CS~29516--024  & 4650 & 1.2 & 1.7 &  --3.06 &--0.63 & --3.55 &  --0.49& --3.67  & 2\\
CS~29518--051  & 5200 & 2.6 & 1.4 &  --2.69 & +0.45 & --2.47 &  +0.22 & --2.82  & 2\\
CS~30325--094  & 4950 & 2.0 & 1.5 &  --3.30 &--2.37 & --5.29 &  --1.99& --      & 2\\
CS~31082--001  & 4825 & 1.5 & 1.8 &  --2.91 & +0.52 & --2.40 &  +0.51 & --2.17  & 4\\
additional giants\\
CS~22891--200  & 4700 & 1.3 & 2.0 &  --3.65 & --1.75& --4.67 &  --1.02& --4.28  & 2\\
CS~22949--048  & 4800 & 1.5 & 2.0 &  --3.25 & --1.39& --4.31 &  --1.06& --4.33  & 2\\
CS~22950--046   & 4650 & 1.0 & 2.2 &  --3.56 & --1.12& --4.04 &  --0.48& --4.49  & 2\\
\hline                                       
\end{tabular}
\end{center} 

\end{table*}

\section{Atomic model and calculations} 
\subsection{The strontium atomic model} 
A NLTE analysis of the \ion{Sr}{ii} spectrum has been the subject of only a few papers. 
A first application of the NLTE calculations for Ap-Bp stars was made by 
Borsenberger et al. (\cite{BOR81}), who used a six-level (plus continuum)  
atomic model of \ion{Sr}{ii}. Belyakova \& Mashonkina (\cite{BM97}) 
considered the statistical equilibrium of \ion{Sr}{ii} in atmospheres of F-G 
stars of different luminosity classes with metallicity ranging from zero 
to --3. They used a model atom consisting of 40 levels  of \ion{Sr}{ii} and 
the ground level of the \ion{Sr}{iii} ion, and this model was applied to derive 
NLTE corrections. After this, Mashonkina \& Gehren (\cite{MG01}) determined 
NLTE abundances in 63 cool stars with metallicity ranging from +0.25 to --2.20.
Mashonkina et al. (\cite{MZG08}) published results for the NLTE abundances of some 
heavy elements (including strontium) in four very metal-poor (VMP) stars with metallicity 
in the range from --2.1 to --2.7.  

\begin{figure}
\resizebox{\hsize}{!}
{\includegraphics{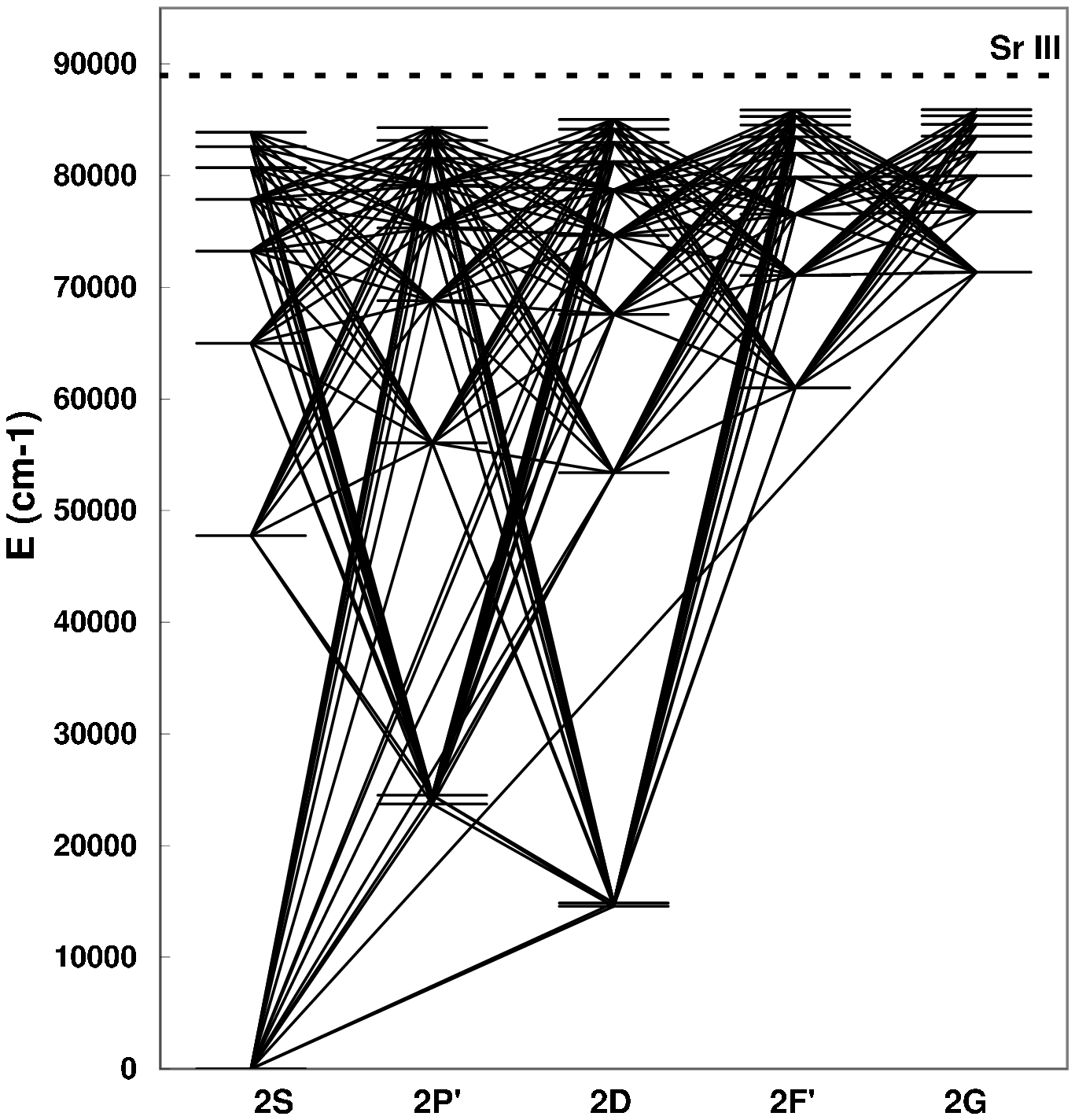}}
\caption[]{The Grotrian diagram for \ion{Sr}{ii} atom.}
\label{GrDiag}
\end{figure}

Our strontium atomic model is quite similar to that first proposed by 
Belyakova \& Mashonkina (\cite{BM97}). The Grotrian diagram is given in Fig.~\ref{GrDiag}. 
The model consists of 44 levels of \ion{Sr}{ii} with $n < 13$ and $l < 6$, as well 
as the ground level of \ion{Sr}{iii} (no excited levels of this ion were 
included, since they have very high excitation potentials). For the two important terms 
$4d{}^{2}D$ and $5p{}^{2}P^{~0}$, the fine structure was taken into account. For the other 
levels, it was ignored.

Since the ionization potential of the neutral strontium is only 5.7~eV, even 
in the atmospheres of cool stars it exists almost only in the form of 
\ion{Sr}{ii}. Therefore, the 24 levels of \ion{Sr}{i} were included in the 
model only for the particle number conservation.  

\subsection{The calculations} 

For all considered levels, the excitation energies were taken from 
Lindgard \& Nielsen (\cite{LN77}). Oscillators strengths for 72 bound-bound 
transitions were selected from three papers: Wiese \& Martin (\cite{WM80}), 
Warner (\cite{WB68}), and Goldsmith \& Boxman  (\cite{GB81}). For the remaining 
258 transitions, we used oscillator strengths from Lingard \& Nielsen 
(\cite{LN77}). 

Radiative photoionization rates for the $s$, $p$, and $d$ levels are based on 
the photoionization cross-sections calculated with the quantum defect method 
and the corresponding tables of Peach (\cite{PG67}). For $f$ and $g$ levels, we 
used the hydrogen-like approximation (Lang \cite{LA98}). 

Collisional rates for the transitions between the ground level $5s{}^{2}S$ and 
$5p{}^{2}P^{~0}$, $5d{}^{2}D$, and $6s{}^{2}S$ levels were estimated with the help 
of the corresponding formula from Sobelman et al. (\cite{SOB81}). For the remaining 
levels, the excitation rates by electron collisions were approximated using 
the formula of van Regemorter (\cite{VRE62}). Collisional excitations for the 
forbidden transitions were taken into account using the semi-empirical formula 
of Allen (\cite{ALL73}) with a factor of 1. Collisional ionization caused by 
electrons was described using the Drawin (\cite{DRA61}) formula. Inelastic 
collisions with hydrogen atoms were taken into account with the  formula of 
Steenbock \& Holweger (\cite{SH92}) with a scaling coefficient 0.01. 
This coefficient was derived from the fitting of the \ion{Sr}{ii} line profiles
in the solar spectrum (Kurucz et al. \cite{KUF84}). 

The NLTE strontium abundance was determined with a modified MULTI code 
(Carlsson \cite{CAR86}). Modifications are described in Korotin et al. 
(\cite{KAL99}). Since we use Kurucz's atmosphere models calculated with ATLAS~9 
(Kurucz \cite{KurCD13}, \cite{K2005}), the necessary background opacities for 
MULTI are also taken from ATLAS~9. In the modified version, the mean intensities 
used to obtain the radiative photoionization rates, are calculated for 
the set of frequencies at each atmospheric layer, and then stored 
in a separate block, where they can be interpolated. This enables us to take 
account of the absorption in a great number of lines, especially in the UV 
region, which is important for the accuracy of the photoionization 
rate calculations.  

After the combined solution of the statistical equilibrium and radiative 
transfer equations, we derived the populations of the $5s{}^{2}S$, $4d{}^{2}D$, 
$5p{}^{2}P^{~0}$, and $6s{}^{2}S$ levels allowing us to reproduce the profiles of 
the lines of interest, provided that the broadening parameters were known. 
Those parameters were taken from the VALData-base 
\footnote{http://ams.astro.univie.ac.at/vald/}.

In the visual part of the stellar spectra and near-IR range, there are only a 
few detectable lines of \ion{Sr}{ii}, including the resonance lines 
4077~\AA, 4215~\AA, and the subordinate lines 10036~\AA, 10327~\AA, and 10914~\AA. 
We note that resonance lines are significantly blended with iron, chromium, as well 
as some molecular lines. The IR lines are not blended. The 10914~\AA~line is 
unavailable in the UVES spectra, but the two other IR lines 10036~\AA~ and 10327~\AA~ 
are, even though in some EMP stars they are both detected at the noise level.

To compare the NLTE strontium line profiles to the observed spectrum in the visual 
part of the spectrum one needs to use a combination of the NLTE and LTE synthetic 
spectrum since the resonance \ion{Sr}{ii} lines  are to some extent blended with 
other metallic lines. This was achieved using the updated code SYNTHV (Tsymbal, 
\cite{TSY96}) which is designed for synthetic spectrum calculations in LTE. With 
this program, we calculated a synthetic spectrum for particular wavelength ranges
including the \ion{Sr}{ii} lines of interest taking into account all the lines from 
each region listed in the VALData-base. For the strontium lines, the 
corresponding $b$-factors (factors of deviation from LTE level populations) that 
had been calculated in MULTI were included in SYNTHV, where they were 
used in calculating the strontium line source function (in particular, the first 
plot in Fig.~\ref{spectra} gives an example of this procedure application). 

\subsection{Test calculations} 

All the above listed input data and our strontium atomic model were checked 
using several test calculations. For this, we used high-resolution 
spectra of some stars with well known parameters.

First of all, we calculated profiles for the six \ion{Sr}{ii} lines in the 
solar spectrum (visual and infrared range, Kurucz's et al. 1984 solar flux 
spectrum) using the Kurucz (1996) solar atmosphere model with 
microturbulence of 1~km~s$^{-1}$. The strontium lines parameters are given 
in Table~\ref{lines}. 

\begin{table}
\begin{center}    
\caption[]{Parameters of the \ion{Sr}{ii} lines.}
\label{lines}
\begin{tabular}{cccccc}
\hline
$\lambda$,\AA &   $f$  & $\log~\gamma_{\rm rad}$ & $\log\gamma_{\rm VW}$ \\
\hline
  4077.7090  & 7.063e-1  &8.130  &  -7.70  \\
  4161.7920  & 1.945e-1  &8.110  &  -7.60  \\
  4215.5190  & 3.303e-1  &8.100  &  -7.70  \\
  4305.4430  & 1.811e-1  &8.080  &  -7.60  \\
 10036.6530  & 1.219e-2  &7.340  &  -7.63  \\
 10327.3110  & 7.393e-2  &7.320  &  -7.63  \\
 10914.8870  & 5.754e-2  &7.270  &  -7.63  \\
\hline
\end{tabular}
\end{center}
\end{table}

\begin{figure}
\resizebox{\hsize}{!}
{\includegraphics {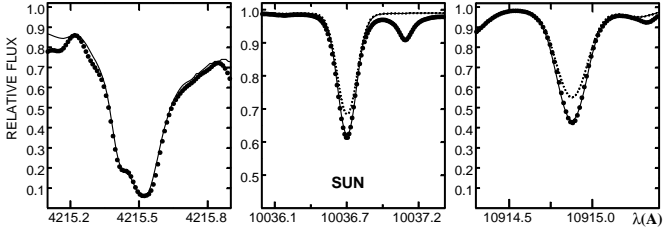}}
\caption[]{Profile fitting of the \ion{Sr}{ii} lines in the solar spectrum. Observed  
spectrum - dots,  NLTE profiles - smooth line, LTE profiles - dotted line.}
\label {spectra}
\end{figure}

The \ion{Sr}{ii} lines in the solar spectrum are closely reproduced by our 
calculations (see Fig.~\ref{spectra}) with strontium abundance (Sr/H) = 2.92 
(as recommended by Asplund et al. \cite{Asp05}). For the sake of completeness, 
we also provide pure LTE profiles for all investigated lines.

We also compared the calculated \ion{Sr}{ii} line profiles with those observed 
in the following well studied stars: Arcturus, Procyon, HD~1581, and the metal-deficient 
star HD~122563 (which is also in our list of program stars). For these stars, spectra 
are available from the UVES archive of the Paranal Observatory Project (Bagnulo et al. 
\cite{BAG03}). These spectra were obtained with R = 80000 and have S/N of more than 300. 
Since the third \ion{Sr}{ii} IR line 10914 \AA~ is unavailable in the UVES spectra (but is 
important for the elaborated strontium atomic model check), we used the Keck IR spectrum 
of HD~122563 previously analyzed in Smith et al. (\cite{SMI04}). Adopted parameters of 
these stars are given in Table~\ref{stars}. The estimated NLTE strontium abundance of
these stars is given in the last column of this table.

\begin{table}
\begin{center}    
\caption[]{Fundamental parameters of the reference stars and their NLTE Sr abundance.}
\label{stars}
\begin{tabular}{ccccccc}
\hline
Star      & T$_{\rm eff}$, K & $\log~g$ & V$_{\rm t}$, km~s$^{-1}$ &  [Fe/H] & (Sr/H) \\
\hline
Arcturus  &   4300          &    1.5   &    \bf{1.7}   & --0.33   &  2.26 \\
Procyon   &   6500          &    4.0   &    \bf{1.8}   &  +0.00   &  3.12 \\ 
HD 1581   &   6000          &    3.9   &    \bf{1.0}   & --0.22   &  2.69 \\ 
HD122563  &   4600          &    1.1   &    \bf{2.0}   & --2.82   &  0.10 \\
\hline
\end{tabular}
\end{center}
\end{table}

\begin{figure}
\resizebox{\hsize}{!}
{\includegraphics {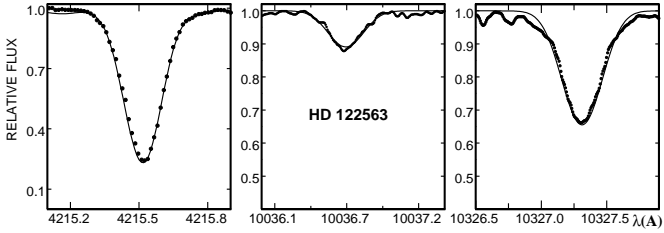}}
\caption[]{The same as Fig.~\ref{spectra} but for HD~122563 spectrum. No LTE profiles 
are shown.}
\label {spectrb}
\end{figure}

Figure~\ref{spectrb} shows that the \ion{Sr}{ii} lines in HD~122563 that belong 
to the different multiplets are well reproduced with a single strontium 
abundance.

Very good agreement is obtained between the abundances derived from the lines 
in the infrared and visible domains for the stars HR~1581 and HD~122563. 
For HD~122563, the abundance of Sr (and Ba) is also in very good 
agreement with the NLTE results of Mashonkina et al. (\cite{MZG08}).

\section{The behaviour of the $b$-factors}
\begin{figure}
\resizebox{\hsize}{!}
{\includegraphics {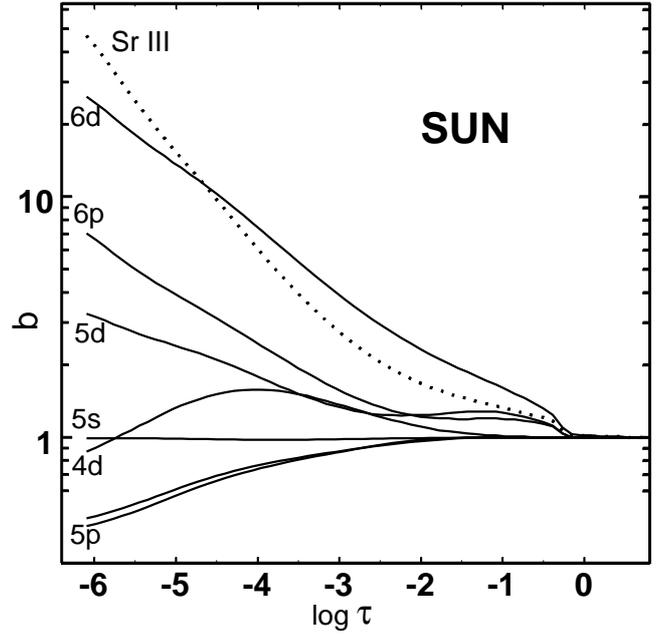}}
\caption[]{The distribution of the $b$-factors in the solar atmosphere.}
\label {bfact}
\end{figure}

Figure~\ref{bfact} shows the behaviour of the $b$-factors for the solar atmosphere.
As one can see, $b$-factors of the $5p$ level are smaller than those for $5s$ level 
over the whole atmosphere. At the same time, one can note that since \ion{Sr}{ii} is 
a dominating ionization stage of strontium at these temperatures, the population of 
the ground $5s$ level does not deviate significantly from an equilibrium population. 
Thus, the resonance lines are stronger, but only slightly because in the range of 
their formation the $b$-factors of the $5s$ level are close to unity.  

For the solar metallicity, the lines 4077 \AA~ and 4215 \AA~ are only slightly 
affected by the NLTE effects (abundance correction is about 0.01), while for 
all three IR lines, corrections are rather significant (about 0.3).   

\begin{figure}
\resizebox{\hsize}{!}
{\includegraphics {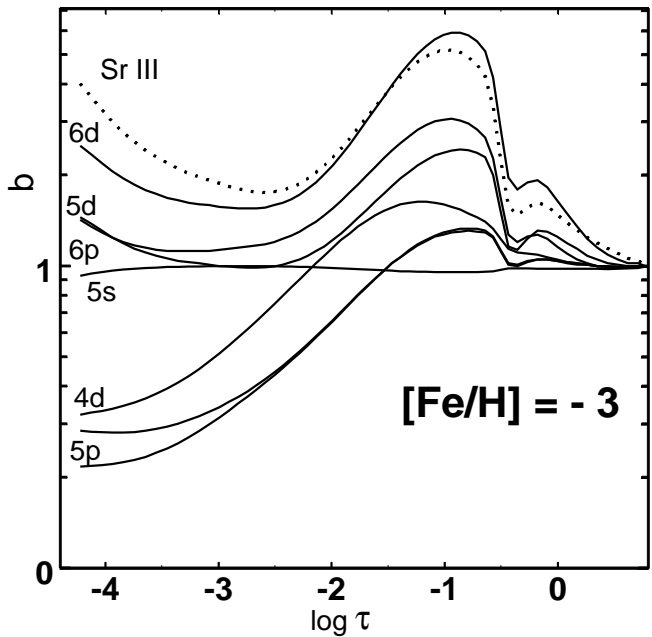}}
\caption[]{The same as Fig.~\ref{bfact} but for the metal-poor atmosphere.}
\label {bfactd}
\end{figure}

For the giant stars with metal-deficient atmospheres ([Fe/H= --3 and 
$T_{\rm eff} = 5000$ \, K), the NLTE correction is somewhat different 
(Fig.~\ref{bfactd}). At low metallicity, all the lines are formed, 
on average, in the deeper atmosphere layers. The lower electron concentration 
leads to significant deviations from LTE in the level populations. 
The ground level of \ion{Sr}{ii} is underpopulated at $\log~\tau = - 1$, 
while excited levels are overpopulated. The pumping mechanism increases 
the population of $5p$ level, and consequently the populations of the higher 
levels become overpopulated too (since $4d$ level is more tightly connected 
to $5p$ than to $5s$, it is also overpopulated). 

For the infrared lines, the $b$-factor of the higher level is smaller than that 
of the lower level and the $b$-factor of the lower level is larger than unity. 
Altogether, this means that IR lines tend to be stronger. The corresponding 
NLTE corrections are about --0.5 dex and even larger depending on the 
metallicity of the atmosphere and other parameters.

\begin{figure}	
\resizebox{\hsize}{!}
{\includegraphics{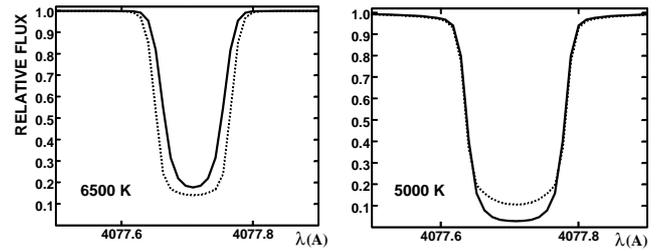}}
\caption[]{LTE  (dotted line) versus NLTE (continuous line) profiles for the two  
temperature values.}
\label {prof}
\end{figure}

The behaviour of the resonance lines were first described in detail by 
Belyakova \& Mashonkina (\cite{BM97}), and later also considered by 
Short \& Hauschildt (\cite{SH06}). Here we mention only two cases, which 
are graphically presented in Fig.~\ref{prof}. At the temperature of 6500~K 
and [Fe/H] lower than --2, these lines becomes weaker than those for LTE, 
and this effect progressively increases with decreasing model metallicity. 
This behavior reflects  that at this temperature and metallicity 
\ion{Sr}{ii} is no longer a completely dominant ionization stage (only up to 
70\% atoms persist in the form of a singly ionized strontium), and the ground 
level of \ion{Sr}{ii} appears to be significantly underpopulated. For the lower 
temperatures, the ratio of \ion{Sr}{iii} to \ion{Sr}{ii} is about 0.01, and 
the ground level tends to have an equilibrium population. The change in the 
profile between NLTE and LTE is small, while the equivalent width is practically 
the same for both LTE and NLTE. This is because the line core is formed in the 
upper atmosphere layers where the $b$-factor for $5p$ level is much smaller 
than for $5s$ (NLTE profile is deeper than LTE), while the far wings are 
formed at the depth where the situation is opposite. This is a result of the 
difference in the flux redistribution between the line core and wings. Since 
in reality the far wings of the line are ignored, the direct use of the equivalent 
width value in NLTE analysis often produces not quite correct results. 
In the plot at the right in Fig.~\ref{prof}, we show the LTE and NLTE profiles 
of the line that comprise almost equal equivalent widths (as follows from the
calculations), while the line core regions look different. Therefore, the 
correct means of deriving NLTE strontium abundance should be individual profile 
fitting for each star using a complete NLTE computation.

\begin{figure}	
\resizebox{\hsize}{!}{\includegraphics{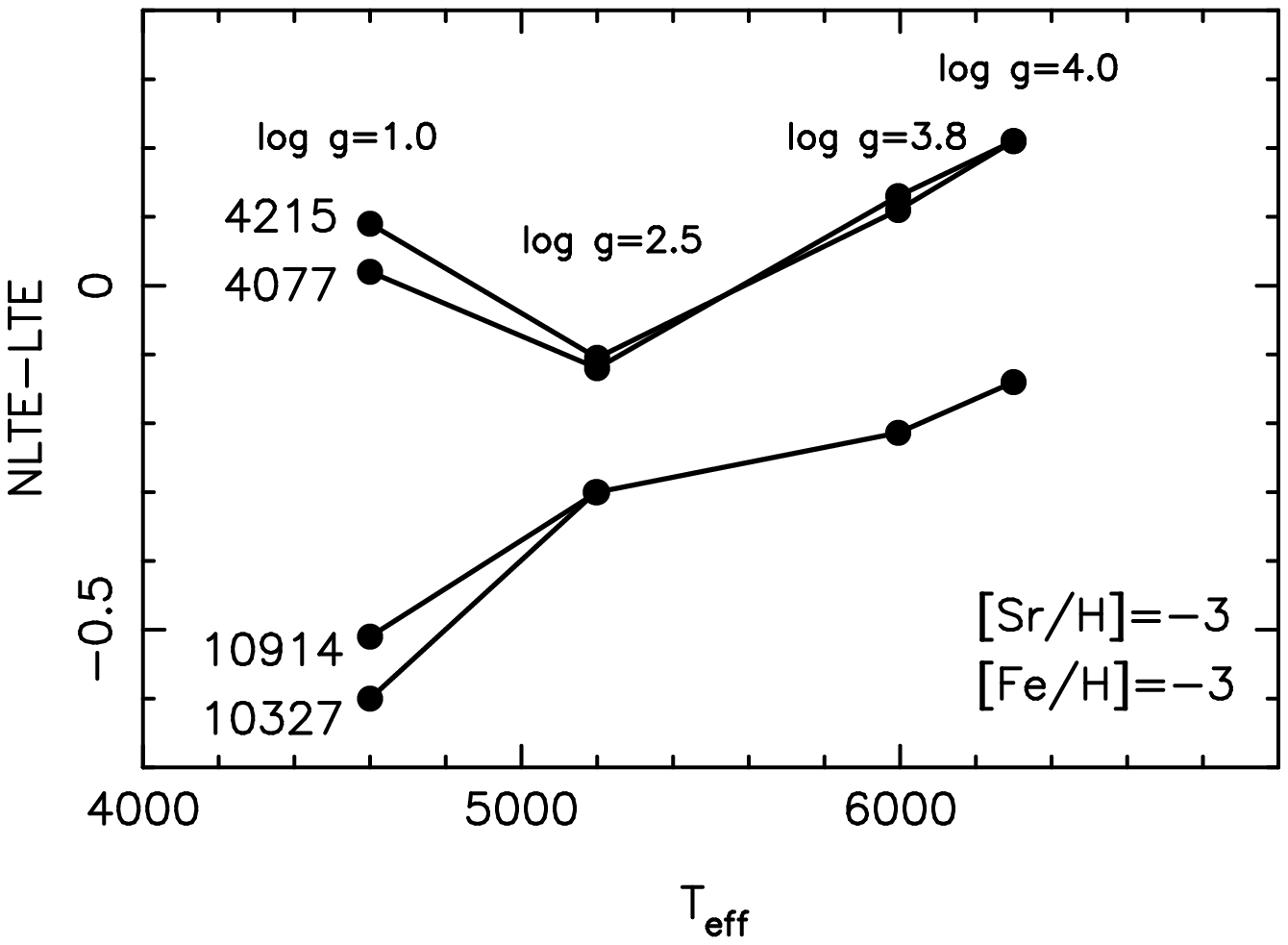}}
\resizebox{\hsize}{!}{\includegraphics{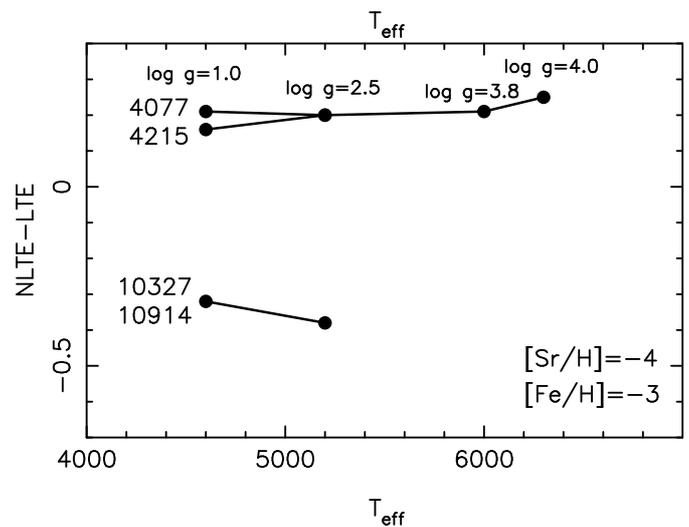}}
\caption[]{NLTE correction as a function of T$_{\rm eff}$ for the different parameters.}
\label {corr}
\end{figure}

In Fig.~\ref{corr}, we show the dependence of  NLTE-LTE abundance correction 
upon effective temperature and other parameters. We note that NLTE corrections 
of the strontium abundance depend on the effective temperature and gravity of 
the model, as well as the strontium abundance itself. 

\section{ NLTE strontium abundance determination in programme stars}
The NLTE abundances of strontium derived in our EMP stars are given in 
Table~\ref{tabstars}, and graphically displayed  in Fig.~8. The NLTE barium 
abundances are also provided in the table (when available); they were
computed with overshooting Kurucz models (Andrievsky et al. \cite{ASK09}) 
and are, to ensure homogeneity with Sr abundances, shifted by --0.03 dex for 
the turn-off stars and --0.05 dex for the giants.

The Sr and Ba abundance results shown in the following diagrams are 
all NLTE abundances.
 
This work confirms (with some improvements) the main trends previously found for the Sr 
abundances, determined assuming LTE, especially those for the same sample of EMP stars: 
Fig.~8, is directly comparable to Fig. 12 (upper panel) of Bonifacio et al. (\cite{BSC09}). 
A comparison of the figures (and of the tables) indicates that the NLTE corrections are 
moderate, but not negligible. The NLTE strontium abundances are, on average, about 
0.2 dex higher than the LTE abundances, with variations from star to star.


It has long been known that the scatter in the Sr abundance is very large 
(e.g. McWilliam, \cite {MW98}). The scatter in the [Sr/Fe] ratios is slightly 
smaller when NLTE  (rather than LTE) values are used, but even using the NLTE values, 
this scatter remains very large (Fig.~8). The scatter is also larger at low metallicities 
(e.g.: McWilliam \cite {MW98}, Fran\c cois et al. \cite{FDH07}, Bonifacio et al. \cite{BSC09}, 
Lai et al. \cite{LAI08}): this trend seems to be confirmed in NLTE (Fig.~8), although 
we note that this conclusion depends on the position of the C-rich star CS~22949-037; 
since this is also an $\alpha$-element-rich star and therefore does not have a very low metallicity 
(Aoki et al. \cite{AHB05}), its extremely low [Fe/H] value might not be representative.   

Despite the large scatter, the abundances of the EMP stars show that, as the 
iron abundance increases, the Sr abundance increases, in the mean, far more than Fe, 
the  [Sr/Fe] ratio increasing rapidly at first (faster than a secondary process) and 
stabilizing at about the solar [Sr/Fe] ratio after some overshoot. The additional   
productions of Sr and Fe then remain at this constant ratio, up to metallicities of
around about [Fe/H] = --1.5, and the scatter decreases (see Fran\c cois et al., \cite{FDH07}).
       
The large enhancement of Sr, already detected for the LTE measurements (McWilliam \cite {MW98}, 
Fran\c cois et al., \cite{FDH07}) in the (Ba-poor) star CS~22897-008 
(diamond at left in Fig.~8) is confirmed by our NLTE determinations. This enhancement is larger 
than for any of those listed  previously in "Sr-rich metal-poor stars" (Truran et al. \cite{T02}).  
The star is intermediate between the Sr-rich stars HD~122563 and the (even richer) star BS~16550-087 
(Lai et al. \cite{LAI08}), but it has a lower metallicity (and  a lower Ba abundance) than the 
two Sr-rich stars. It could be, as well as  HD~122563, considered as an example of a so-called  
"weak" r-process (Honda et al. \cite{HA07}): its Eu abundance is too low to be measured but, 
by that very fact, could be compatible with such a "weak" r-process. 

The low metallicity turn-off star CS 29527-015 (extreme left closed circle
in Fig.~8) is Sr-rich, its Ba and Eu abundances being too low to be measured.

\begin{figure}	
\resizebox{\hsize}{!}
{\includegraphics{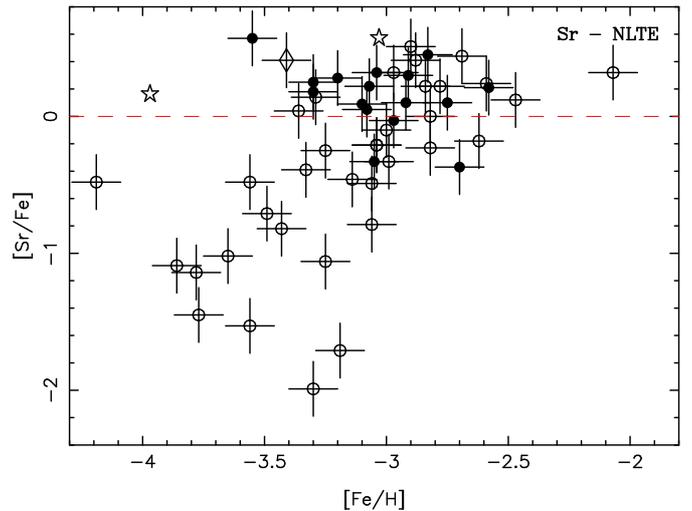}}  
\caption[]{ [Sr/Fe] vs. [Fe/H] in our program stars: dwarfs: $closed~circles$, giants: 
$open~circles$, the  C-rich stars CS~22949-037 and CS~22892-052: $asterisks$, 
the Ba-poor star CS~22897-008: $diamond$. Three newly analyzed giants are added.
Typical error bars have been indicated.}
\label {srfeH}
\end{figure}

\subsection{Sr vs. Ba}
Fig.~9 shows [Sr/Fe] versus [Ba/Fe]. Both Sr and Ba globally increase with Fe. 
Since the Sr/Fe ratio has a larger scatter at low metallicities, it also has
more scatter at low values of [Ba/Fe].

In this figure, the thick line corresponds to the constant  ratio [Sr/Ba] = -- 0.31, 
which is the value chosen by Qian \&  Wasserburg (\cite{QW08}) to represent their 
component for the Sr-poor stars. This ratio is compatible with the production ratio 
of Sr and Ba for the main r-process: it is near (but very slightly smaller than) the ratio 
given by Burris et al. (\cite{BPA00}) for the  r-process-only component of these elements 
in the Solar system, it is also slightly below the similar value  provided for this component  
by Arlandini et al. (\cite{Arl99}).

In Fig.~9, all the stars (apart from one, marginally) are situated higher than the line 
suggesting that, for most stars, one (or more) additional production(s) of Sr occurred 
(or a variation in the main r-process, or both). The departures from the main r-process 
are higher at low metallicities (and at low Ba abundances).

In our sample, two r-rich stars (CS~31082-001,  CS~22892-052) defining the main 
r-process (Cowan \& Sneden, \cite {CS06}), are (as expected) near the solid line 
(and also BD~+17$\degr$3248). In their survey of r-rich stars (r-II stars), Barklem et al. 
(\cite{BAR}) note that the r-II stars are centred on a metallicity of [ÊFe/HÊ] = --2.81,  
and the two r-rich giants in our sample, have indeed such relatively high metallicities 
(and are consequently relatively  Sr and Ba-rich, in the figure being near the upper corner). 
The Sr-rich turn-off star (closed circle at left in Fig.~\ref{srfeH}) has a significanly lower 
metallicity.

In Fig.~9, it is seen that the upper envelope has a [Sr/Fe] ratio that is approximately constant, 
while [Ba/Fe] increases by an order of magnitude, suggesting that the production process of Sr 
is different (somehow independent?) of the process producing Ba. With increasing Fe, [Ba/Fe] 
increases thanks to the main r-process, while along the upper envelope [Sr/Fe] does not. 
This at least indicates that an early production of Sr occurred in the chemical evolution of the Galaxy.


\begin{figure}	
\resizebox{\hsize}{!}
{\includegraphics{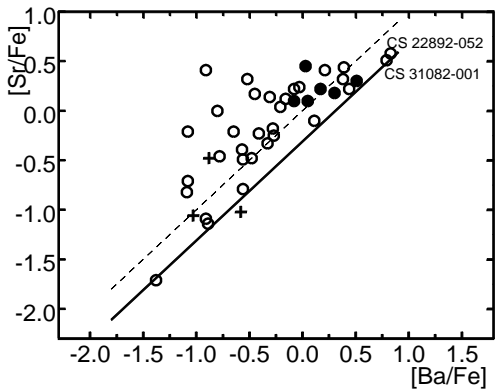}}
\caption[]{[Sr/Fe]  vs. [Ba/Fe].  The solar ratio [Sr/Ba] = 0 is represented by 
the $dashed~line$. The  value [Sr/Ba] = --0.31, corresponding to the main r-process 
production, is represented by the $full~line$. $Closed~circles$: dwarfs, $open~circles$: 
giants, $crosses$ : three added EMP giants. The stars without Ba measurements are not 
represented. Symbols as in Fig.~8. In Fig.~9 and in following figures, the error bars 
are no longer indicated for an easier visual perception of the abundance trends.}
\label{srba}
\end{figure}


Fig.~10 shows the positions of the stars in the  [Sr/Ba] vs. [Fe/H] plot, together with 
a plot of the theoretical LEPP model predictions (Travaglio et al. \cite{TGA04}). The 
corresponding theoretical curve is not far from  the upper envelope of the data points, 
and an inhomogeneous mixing  of the products of such a LEPP process with the products 
of the main r-process could explain the distribution of the stars in this figure.
 
In our sample, the carbon-rich EMP star CS~22949-037 is classified 
as a "CEMP-no" star: this classification is explained e.g. by Sivarani et 
al. (\cite {SBB}). This carbon-rich star with the rather high ratio [Sr/Ba] $\approx 0.6$
could also be explained by the LEPP process (or by massive rotating stars e.g., Pignatari 
et al. \cite{PGM}, Meynet et al. \cite{MHE10} or other processes; see discussion below). 

 

\begin{figure}	
\resizebox{\hsize}{!} {\includegraphics{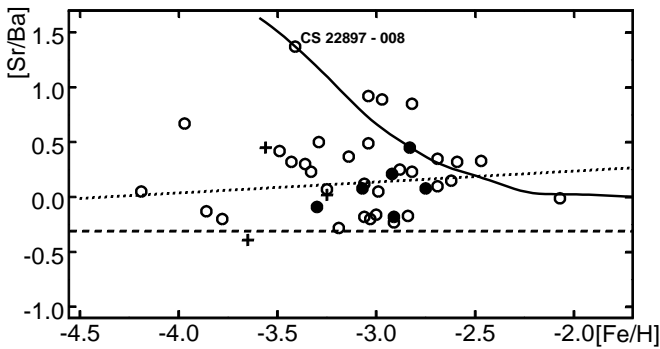}}
\caption[]{[Sr/Ba]  vs. [Fe/H]. LEPP model prediction represented by the 
$thick~line$, [Sr/Ba] = --0.31 : $dashed~line$, linear regression : 
$dotted~line$.}
\label {srbafe}
\end{figure}

\subsection{Production of Fe, Sr, and Ba} 

Fran\c cois et al. (\cite{FDH07}) demonstrated (their Fig. 15) that in extremely metal-poor 
stars, the ratio [Sr/Ba] (as well as Y/Ba and Zr/Ba) is often higher than solar in the range 
--3.7 $>$ [Ba/H] $>$ --4.7. A correlation is found between [Sr/Ba] and [Ba/H] in the interval  
--5.5 $<$[Ba/H]$<$ --4.5 (LTE values). At very low metallicity, Fran\c cois et al. (\cite{FDH07}) 
find that, for the lower observed values of [Ba/H] (about --5 dex), the ratio [Sr/Ba] 
returns to the solar values. Our NLTE homogeneous determinations qualitatively confirm 
this behaviour, and enable one to robustly claim that the Sr abundances are generally higher 
than those predicted by the main r-process pattern. 

Fig.~11 shows that, with [Ba/H] increasing from --5.0 to --4.0, the upper envelope of the scatter 
diagram reaches the one to one slope when excluding the star CS 22897-008 (and even more when 
including this star). Along this envelope, the Sr abundance increases much more than the Ba abundance.  
In the other part of the figure, where [Ba/H]  increases from --3.5 to  --2.5, the upper envelope 
decreases with a slope around --0.7; the Sr abundance increases more gradually than Ba. This behaviour 
illustrates the well-known lag between the onsets of Sr and Ba production, again suggesting that 
the production processes of Sr and Ba are different.

Many stars are relatively near the dashed line (corresponding to the main r-process), about half of 
the stars having Sr/Ba ratios smaller than three times the main r-process ratio. 
 
The three stars with the lowest  Ba abundances  (including CD~-38$\degr$245) have Sr/Ba ratios near 
the main r-process ratio (in agreement with the behaviour of the most Ba deficient star of 
Lai et al. (\cite{LAI08}).

\begin{figure}	
\resizebox{\hsize}{!}
{\includegraphics{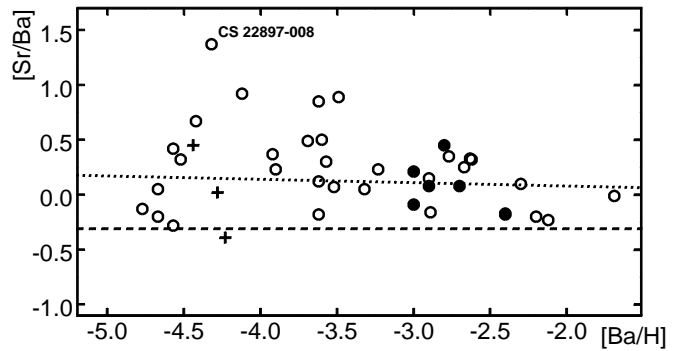}}
\caption[]{[Sr/Ba] vs. [Ba/H]. [Sr/Ba]$_{H}$ = --0.31 - $dashed~line$, linear 
regression - $dotted~line$.}
\label {srbaba}
\end{figure}
 
\begin{figure}	
\resizebox{\hsize}{!}
{\includegraphics{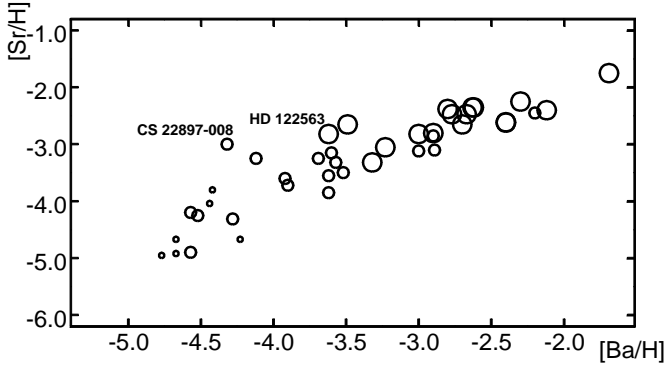}}
\caption[]{[Sr/H] vs. [Ba/H]. Stars with [Fe/H] $< -3.5$ - $small~circles$,
$-3.5<$ [Fe/H] $<-3.0$ -$intermediate~circles$, [Fe/H]$>-3.0$ - $large~circles$. }
\label {srba1}
\end{figure}

Fig.~12 shows the simultaneous enrichment of Fe, Sr, and Ba, the low metallicity stars 
being both Ba-poor and Sr-poor. Fig.~\ref{srba1} 
is somewhat similar to 
Fig.~9 but provides an indication of the iron production. 

Summarizing, we can state that our NLTE Sr and Ba abundances discussed above appear to be 
consistent with the Qian \& Wasserburg (\cite{QW08}) simple three-component model of chemical
evolution. According to that model, the first phase of the chemical evolution of the Galaxy 
would be the production of elements by high-energy hypernovae: their yields would include 
significant amount of low-A elements (from Na to Zn, including iron), but no heavier species. 
In reality, Fig.~12 shows that iron-poor stars have a very low content of Sr and Ba. The shortfall 
in Sr later decreases to zero when the low-mass and normal-mass SNe ($H$ and $L$ sources respectively) 
begin producing strontium (Fig.~ 8). Moreover, since Sr is produced by both the $H$ and $L$ sources, 
and Ba is produced predominantly by the lower-mass $H$ source, it can explain the general
decrease in the Sr/Ba scatter, which is clearly seen seen in Fig.~11. 


\subsection{Comparison to faint dwarf spheroidal galaxies}

We have much to learn about dSphs, but a full discussion is beyond the scope of this paper. 
We present only two short comparisons.

Cohen \& Huang (\cite{CH10}) find, in the (metal-poor) stars of the Ursa Minor dwarf 
spheroidal galaxy, the heavy elements produced follow the main r-process pattern: this 
is consistent with the short duration of the star-forming phase of this small 
galaxy, inferred from the CMDs  (Orban et al. \cite{Orb08}). 

Norris et al. (\cite{NGW10}) identify CEMP-no stars (or similar) in dwarf 
galaxies, and note a similarity between their abundance patterns and those of 
the same stars in our Galaxy: the similarity is indeed striking, but strontium 
is systematically stronger in the stars of our Galaxy. The comparison is made 
for LTE abundances, but for one of them (CS 22949-037), the high Sr 
abundance is confirmed by our NLTE determinations.

\subsection{Sr formed by other processes?}

Many independent explanations of excess of Sr production relative to the main
r-process have been proposed, such as 

1) a quenched flux model (Truran et al. \cite{T02});

2) particular models of hypernovae (e.g. special computation of four models of hypernovae 
reproducing the high Sr/Ba ratios (and other abundances) in four stars of our sample, by 
Izutani et al. (\cite{IUT09});
 
3) the inhomogenous mixing of three types of nucleosynthesis productions:
a phenomenological "three-component model" ( Qian \& Wasserburg \cite{QW08});
 
4) a weak r-process (Honda et al.  \cite{HA07}) or a variation in the 
r-process  (Roederer et al. \cite{RCK10});
 
5)  contamination by main s-process (by an AGB companion): 
metal-poor  AGB  do not generally provide a strong Sr  excess, still, 
monitoring a subsample of Sr-rich stars would be useful;

6) a LEPP process (Travaglio et al. \cite{TGA04}). 

Since nitrogen  is observed as a primary product in some EMP stars (Spite et al. \cite{SM2005}),  
rotating massive stars may produce heavy elements, and in some cases 
strong Sr/Ba ratios (Pignatari et al. \cite{PGM}). However, the correlation 
of Sr/Ba with the abundance of N is weak, suggesting that, if rotating stars are 
a source of Sr, it is not the unique additional source. The Sr-rich star CS~22949-037, 
classified as CEMP-no, might be produced by a LEPP, or by a massive rotating star (producing 
also carbon).

\section{Conclusion}
The homogeneous NLTE Sr abundances  of the stars of this sample, provide  
precise information about the Sr abundances in the early evolution of the Galaxy. 
In addition to previously determined NLTE Ba abundances, these data provide some 
insight into the process(es) of Sr production. 

The methods of the NLTE determination of Sr abundances have been verified (solar 
spectrum,  extension in infrared), the results being in satisfactory agreement 
with computations found in the literature (e. g. Belyakova \& Mashonkina \cite{BM97},
Mashonkina et al. \cite{MZG08}) and  comparisons with a few stellar 
NLTE abundances available in the literature show good agreement.

Our study here has confirmed the behaviour of the Sr LTE abundances  found  in the 
literature for both metal-poor stars (e.g. by McWilliam, \cite{MW98} and Lai et al.
\cite{LAI08}) and our sample (Fran\c cois et al. \cite{FDH07} 
and Bonifacio et al. \cite{BSC09}). These behaviours are now established on 
a firmer basis. 

During the chemical evolution of the Galaxy, when the iron abundance increases, 
the mean Sr abundance increases, the mean [Sr/Fe] ratio increases, reaching 
(and then exceeding) the solar ratio, before  decreasing slowly towards the solar 
ratio.

The Sr/Ba ratio is always higher than the main r-process ratio. In addition to the 
large scatter in the Sr/Ba ratios, this suggests that either some production 
of Sr (at least partly independent of the production of Ba) is added 
to the main r-process, or that the r-process that produces Sr and Ba varies 
significantly.
 

A number of additional processes of Sr enrichment have been proposed in the 
literature, although none have been clearly verified, the data even being consistent 
with several processes existing simultaneously, enhancing the difficulty of identifying 
these processes.

In this complex situation, it is of the upmost importance to collect new 
accurate abundances for a larger sample of stars, and a larger number of 
heavy elements (first and second peak elements) in the metallicity range 
--4.5 $<$ [Fe/H] $<$ --2.5, to try to identify the processes 
(and their sites) responsible for producing these elements.

\begin{acknowledgements}

Authors would like to thank Andrea Dupree for sending the Keck IR spectra of two stars: 
HD~1581 and HD~122563. This work made use of the CDS, ADS and VALData-base. 
SMA and SAK are thankful to GEPI for the hospitality during the stay in 
Paris-Meudon Observatoire, and would like to acknowledge partial support from SCOPES grant No. 
IZ73Z0-128180/1 and from the contract UKR CDIV N¡24008 in the France (CNRS) - Ukraine 
(National Academy of Sciences) exchange programmme. FS and MS acknowledge partial support from 
PNPS (CNRS). 
We thank the anonymous referee for valuable comments that improved the paper, 
and language editor Claire Halliday for her help.

\end{acknowledgements}

\end{document}